\documentstyle[aps]{revtex}
\begin{document}
\draft
\title{The Constraint of a General Effective Potential \\
in Vector Torsion Coupled Conformally Induced Gravity}
\author{Jewan Kim$^{*}$, C. J. Park$^{\dagger}$ and Yongsung Yoon$^{\#}$}
\address{$^{*}$Department of Physics, Seoul National University
Seoul 151-742, Korea \\
$^{\dagger}$Institute for Mathematical Science, Yonsei University
Seoul 120-749, Korea \\
$^{\#}$Department of Physics, Hanyang University
Seoul 133-791, Korea
}
\maketitle
\begin{abstract}
It is found that the deviation of an effective potential from the quartic form
is related to the metric and vector torsion dependencies of the effective
potential in the vector torsion coupled conformally induced gravity.

\end{abstract}

\section{Introduction}

It is considerable that gravity is
characterized by a dimensionless coupling constant $\xi$ and
that the gravitational constant $G_{N}$ is given by the inverse square of
the vacuum expectation value of the dilaton field \cite{Zee,Smolin}.
The weakness of gravity can be associated with spontaneous symmetry breaking
at very high energy scale.
The induced gravity action in Riemann space is given by
\begin{equation}
S_{eff}(\phi;g_{\beta\gamma})=\int d^{4}x
\sqrt{g}\{-\frac{1}{2}\xi\phi^{2}R(\{\})
+\frac{1}{2}\partial_{\mu}\phi\partial^{\mu}\phi
-V(\phi;g_{\beta\gamma})\}~.
\label{r_action}
\end{equation}
The effective potential $V_{eff}(\phi;g_{\beta\gamma})$ for the
dilaton field may be
attributed to quantum fluctuations of the conformal factor or gauge fields
coupled to the dilaton field \cite{Percacci,Odintsov,Shapiro,Book}.
In any case, it is assumed that the effective potential attains its minimum
value at $\phi = \sigma$, then the above induced gravity action is reduced to
the well known Einstein-Hilbert action with the gravitational constant,
\begin{equation}
G_{N}=\frac{1}{8\pi\xi\sigma^{2}}~.
\label{grv}
\end{equation}

     On the analogy of the $SU(2) \times U(1)$ symmetry of the weak
interactions, we can consider a continuous symmetry which is broken through
a spontaneous symmetry breaking in the gravitational interactions.
The most attractive symmetry is the local conformal symmetry which
rejects the Einstein-Hilbert action, but
admits the induced gravity action (\ref{r_action})
with the specific coupling $\xi = -\frac{1}{6}$ and the bare quartic potential.
If we extend Riemann space to the minimal Riemann-Cartan space which has
almost Riemannian structure with the additional vector torsion like a conformal
gauge field \cite{Nieh},
the conformally induced gravity action can be written as follows \cite{Smolin};
\[
S_{eff}(\phi;S_{\alpha},g_{\beta\gamma})=\int d^{4}x\sqrt{g}
\{-\frac{\xi}{2}R(\{\})\phi^{2}
+\frac{1}{2}\partial_{\alpha}\phi\partial^{\alpha}\phi
\]
\begin{equation}
-\frac{1}{4}H_{\alpha\beta}H^{\alpha\beta}
+\frac{1}{2}(1+6\xi)S^{\alpha}\partial_{\alpha}\phi^{2}
+\frac{1}{2}(1+6\xi)S_{\alpha}S^{\alpha}\phi^{2}
-V_{eff}(\phi;S_{\alpha},g_{\beta\gamma})\}~,
\label{rc_action}
\end{equation}
where $S_{\mu}=\frac{1}{3}T^{\nu}_{~\nu\mu}$ is the vector torsion
and $H_{\mu\nu}=\partial_{\mu}S_{\nu}-\partial_{\nu}S_{\mu}$
is the conformal field strength.
This action (\ref{rc_action}) is conformally invariant for arbitrary
$\xi$ with the bare quartic potential $\frac{\lambda}{4!}\phi^{4}$.
However, the effective potential $V_{eff}(\phi;S_{\alpha},g_{\beta\gamma})$
radiatively corrected by
quantum fluctuations depends on the metric in general and breaks the local
conformal symmetry. The radiative corrections of the bare quartic potential
can give non-trivial minima and non-vanishing vacuum expectation value $\sigma$
to the dilaton field.
This symmetry breaking in conformally induced gravity might be applied to the
induced gravity inflationary models \cite{La,Turner,Kaiser}.

The quantum effects of the conformal factor for the dilaton field
in conformally non-invariant induced gravity with torsion background
have been considered in Ref.\cite{Antoniadis,Elizalde,Buchbinder,Bytsenko}.
In the symmetry broken phase, the torsion vector gets the mass
$\sqrt{(1+6\xi)}\sigma$, and
the ratio of vector torsion mass $M_{s}$ and
planck mass $M_{p}$ is
\begin{equation}
\frac{M_{s}}{M_{p}}=
\sqrt{\frac{(1+6\xi)}{8\pi\xi}},
\label{tpm1}
\end{equation}
which is independent of
the vacuum expectation value $\sigma$ of the dilaton field.
Therefore, if $\sqrt{(1+6\xi)}$ is small enough,
it can be considered an approximate quantum gravity such that we treat
the metric as a classical background and
the vector torsion as a quantum field in the below plank energy scale
\cite{Ours}.

\section{Equations of Motion for Conformally Induced Gravity}

In this section, we analyze the classical equations of motion for the action
Eq.(\ref{rc_action}).
Varying the action, we obtain the following three equations of motion;
\begin{equation}
\Box\phi=
-\xi R(\{\})\phi
+(1+6\xi)\phi(S^{\mu}S_{\mu}-\nabla_{\mu}S^{\mu})
-\frac{\partial V_{eff}(\phi;S_{\alpha},g_{\beta\gamma})}{\partial\phi}~,
\label{box}
\end{equation}
\begin{equation}
\partial_{\mu}(\sqrt{g}H^{\mu\nu})=
-(1+6\xi)\sqrt{g}\{(\partial^{\nu}\phi)\phi+S^{\nu}\phi^{2}\}
+\frac{\partial V_{eff}(\phi;S_{\alpha},g_{\beta\gamma})}{S_{\nu}}~,
\label{box1}
\end{equation}
\[
\xi\phi^{2}G_{\mu\nu}=
-(H_{\mu\alpha}H_{\nu}^{~\alpha}
  -\frac{1}{4}g_{\mu\nu}H_{\alpha\beta}H^{\alpha\beta})
+(\partial_{\mu}\phi\partial_{\nu}\phi
  -\frac{1}{2} g_{\mu\nu}\partial_{\alpha}\phi\partial^{\alpha}\phi)
+(1+6\xi)\phi^{2}(S_{\mu}S_{\nu}-\frac{1}{2}g_{\mu\nu}S_{\alpha}S^{\alpha})
\]
\[
+(1+6\xi)(S_{\mu}\phi\partial_{\nu}\phi+S_{\nu}\phi\partial_{\mu}\phi
-g_{\mu\nu}S^{\alpha}\phi\partial_{\alpha}\phi)
+\xi\{\nabla_{\mu}(\phi\partial_{\nu}\phi)+\nabla_{\nu}(\phi\partial_{\mu}\phi)
-g_{\mu\nu}\Box\phi^{2}\}
\]
\begin{equation}
+g_{\mu\nu}V_{eff}(\phi;S_{\alpha},g_{\beta\gamma})
-2\frac{\partial V_{eff}(\phi;S_{\alpha},g_{\beta\gamma})}
{\partial g^{\mu\nu}}~.
\label{long}
\end{equation}
Taking the divergence of Eq.(\ref{box1}), we obtain
\begin{equation}
(1+6\xi)\nabla_{\mu}(S^{\mu}\phi^{2})= -\frac{1}{2}(1+6\xi)\Box\phi^{2}
+\nabla_{\nu}\frac{\partial V_{eff}(\phi;S_{\alpha},g_{\beta\gamma})}
{\partial S_{\nu}}~.
\label{tran}
\end{equation}
The trace of Einstein Eq.(\ref{long}) is
\begin{equation}
-\xi R(\{\})\phi^{2}=
-\partial_{\alpha}\phi\partial^{\alpha}\phi
-(1+6\xi)(S^{\alpha}\partial_{\alpha}\phi^{2}+S_{\alpha}S^{\alpha}\phi^{2})
-3\xi\Box\phi^{2}+4V_{eff}(\phi;S_{\alpha},g_{\beta\gamma})
-2\frac{\partial V_{eff}(\phi;S_{\alpha},g_{\beta\gamma})}
{\partial g^{\mu\nu}}g^{\mu\nu}~.
\label{tra}
\end{equation}
{}From Eqs.(\ref{box}) and (\ref{tra}), we have
\[
\phi\Box\phi+\partial_{\alpha}\phi\partial^{\alpha}\phi
+(1+6\xi)\nabla_{\alpha}(S^{\alpha}\phi^{2})+3\xi\Box\phi^{2}=
\]
\begin{equation}
4V_{eff}(\phi;S_{\alpha},g_{\beta\gamma})
-\phi\frac{\partial V_{eff}(\phi;S_{\alpha},g_{\beta\gamma})}{\partial\phi}
-2\frac{\partial V_{eff}(\phi;S_{\alpha},g_{\beta\gamma})}
{\partial g^{\mu\nu}}g^{\mu\nu}~.
\label{combi}
\end{equation}
Using Eq.(\ref{tran}), we have a $\xi$ independent equation for a general
effective potential from Eq.(\ref{combi}) as follows;
\begin{equation}
4V_{eff}(\phi;S_{\alpha},g_{\beta\gamma})
-\phi\frac{\partial V_{eff}(\phi;S_{\alpha},g_{\beta\gamma})}{\partial\phi}
=2\frac{\partial V_{eff}(\phi;S_{\alpha},g_{\beta\gamma})}
{\partial g^{\mu\nu}}g^{\mu\nu}
+\nabla_{\nu}\frac{\partial V_{eff}(\phi;S_{\alpha},g_{\beta\gamma})}
{\partial S_{\nu}}~.
\label{combix}
\end{equation}
Therefore the metric and vector torsion dependencies of an effective
potential are directly related
to the deviation of the effective potential from the quartic form.
Needless to say, if we ignore the metric and vector torsion dependencies
of the effective potential,
then we encounter inconsistency of the classical equations of motion in case
of the vector torsion coupled induced gravity.
Because the relation (\ref{combix}) also appears in case of the special
coupling $\xi = -1/6$, where the vector torsion is decoupled from the dilaton,
we can say that the Eq.(\ref{combix}) generally appears in conformally induced
gravity model.

The equation (\ref{combix}) also requires that the metric independent bare
effective potential should be quartic in the dilaton field.
Because this constraint comes from the assumption that
the bare action is conformally invariant except the potential term,
if we consider non-conformal coupling in kinetic and interacting terms,
there would be no such a constraint.

It would be interesting if it can be explicitly realized that
the effective potential including the quantum fluctuations of
the vector torsion and the dilaton fields with non-vanishing vector torsion
background satisfies the Eq.(11) and shows a relevant phase transition
in general background space-time leading to an inflationary phase
through the conformal symmetry breaking. \vspace{3mm}

\noindent
{\it Acknowledgments:}
This work was supported in part by the Korea Science and Engineering Foundation
and the Ministry of Education through BSRI-2441

\end{document}